\documentclass[11pt]{article}
\usepackage{amsfonts}
\usepackage[margin=2cm]{geometry}
\usepackage{amsmath,amssymb,amsthm,array,booktabs,cite,dsfont,graphicx,mathtools,multirow,rotating,stmaryrd,tensor,verbatim}
\usepackage[bookmarksnumbered,linktocpage,pdfstartview=FitH]{hyperref}
\usepackage[all]{hypcap}
\usepackage[margin=2cm]{geometry}

\graphicspath{{figures/}}

\newcolumntype{M}[1]{>{$}{#1}<{$}}
\newcolumntype{C}[1]{>{\centering}m{#1}}

\newcommand{\rep}[1]{\ensuremath{\mathbf{#1}}}
\newcolumntype{B}[1]{>{\mathbf\bgroup}{#1}<{\egroup}}
\newcolumntype{K}{>{\lvert}{c}<{\rangle}}

\newtheorem*{theorem*}{Theorem}

\numberwithin{equation}{section}
\DeclareMathOperator{\Aut}{Aut}
\DeclareMathOperator{\Str}{Str}
\DeclareMathOperator{\tr}{tr}

\DeclareMathOperator{\Iso}{Iso}

\DeclareMathOperator{\SO}{SO}

\DeclareMathOperator{\SL}{SL}
\DeclareMathOperator{\SU}{SU}
\DeclareMathOperator{\Sp}{Sp}

\DeclareMathOperator{\GL}{GL}
\DeclareMathOperator{\sgn}{sgn}

\newcommand{\be}{\begin{equation}}
\newcommand{\ee}{\end{equation}}
\newcommand{\bea}{\begin{eqnarray}}
\newcommand{\eea}{\end{eqnarray}}

\newcommand{\JOs}{\mathfrak{J}^{\mathds{O}^s}_{3}}

\newcommand{\F}{\mathds{F}}
\newcommand{\R}{\mathds{R}}
\newcommand{\C}{\mathds{C}}
\newcommand{\Q}{\mathds{H}}
\newcommand{\Z}{\mathds{Z}}
\newcommand{\Oct}{\mathds{O}}

\newcommand{\FTS}{\mathfrak{F}}
\newcommand{\FOs}{\mathfrak{F}^{\mathds{O}^s}}

\begin{document}

\begin{titlepage}
\begin{center}
\hfill Imperial/TP/2012/mjd/05\\
\hfill CERN-PH-TH/2012-336 \\

\vskip 1.5cm

{\Huge \bf Freudenthal Dual Lagrangians}

\vskip 1.5cm

{\bf L.~Borsten\,$^1$, M.~J.~Duff\,$^1$, S.~Ferrara\,$^{2,3}$ and
A.~Marrani\,$^4$}

\vskip 15pt

{\it $^1$ Theoretical Physics, Blackett Laboratory, Imperial College London,\\
 London SW7 2AZ, United Kingdom}\\\vskip 5pt
\texttt{leron.borsten@imperial.ac.uk}\\
\texttt{m.duff@imperial.ac.uk}

     \vspace{8pt}

{\it ${}^2$ Physics Department, Theory Unit, CERN,\\
     CH -1211, Geneva 23, Switzerland}\\\vskip 5pt
     \texttt{sergio.ferrara@cern.ch}

     \vspace{8pt}

     {\it ${}^3$ INFN - Laboratori Nazionali di Frascati,\\
     Via Enrico Fermi 40, I-00044 Frascati, Italy}

     \vspace{8pt}

     {\it ${}^4$  Instituut voor Theoretische Fysica, KU Leuven,\\
Celestijnenlaan 200D, B-3001 Leuven, Belgium}\\\vskip 5pt
     \texttt{alessio.marrani@fys.kuleuven.be}

\end{center}

\vskip 2.2cm

\begin{center} {\bf ABSTRACT}\\[3ex]\end{center}

The global U-dualities of extended supergravity  have played a central role in differentiating  the distinct classes of extremal black hole solutions.   When the U-duality group satisfies certain algebraic conditions, as is the case for a broad class of supergravities,  the extremal black holes enjoy a further symmetry known as  Freudenthal duality  (F-duality), which although distinct from U-duality preserves the Bekenstein-Hawking entropy.   Here it is  shown that, by  adopting the doubled Lagrangian formalism,  F-duality,  defined on the doubled field strengths, is not only a symmetry of the black hole solutions, but also of the equations of motion themselves.  A further role for F-duality is introduced in the context of world-sheet actions.  The Nambu-Goto world-sheet action in any $(t,s)$ signature spacetime can be written in terms of the F-dual. The corresponding field equations and Bianchi identities are then related by F-duality allowing for an F-dual formulation of Gaillard-Zumino duality on the world-sheet. An equivalent polynomial ``Polyakov-type'' action is introduced using the so-called \emph{black hole potential}. Such a construction allows for actions invariant under all groups of type $E_7$, including $E_7$ itself, although in this case the stringy interpretation is less clear.




\vfill


\end{titlepage}

\newpage \setcounter{page}{1} \numberwithin{equation}{section}

\section{Introduction}

Much progress has been made in understanding the black hole solutions of supergravity. This line of enquiry has fed into many important insights, perhaps most notably the   microscopic  derivation  of the Bekenstein-Hawking entropy in string theory \cite{Strominger:1996sh}.

A ubiquitous feature of these developments has been the use of symmetry. In particular, extended supergravity theories are furnished with `large' non-compact global symmetries, commonly referred to as U-dualities. The basic structure of such theories can be  understood in terms of these U-dualities. Indeed, when $\mathcal{N}\geq 5$, for which no matter coupling is allowed, the Lagrangian could be considered to be fixed uniquely by the U-duality group, although this logic is perhaps backwards with respect to the usual perspective. It is  not surprising then that U-duality can be used to understand important features of the supergravity black hole solutions. For example, the extremal stationary, single- and multi-centre, supersymmetric and non-supersymmetric solutions can be largely resolved through the group theoretic properties    of a coset $G/H$, where $G$ is the U-duality group of the three-dimensional supergravity obtained by performing a timelike reduction  \cite{Breitenlohner:1987dg,Gunaydin:2000xr,Gunaydin:2005zz,Gunaydin:2005gd,Pioline:2005vi,Gunaydin:2005mx,Gunaydin:2007bg,Gunaydin:2007qq, Gaiotto:2007ag, Bossard:2009at,Bossard:2009mz,Fre:2011aa,Bossard:2011kz}. Another essential component of this story is the \emph{attractor mechanism} \cite{Ferrara:1995ih,Strominger:1996kf,Ferrara:1996dd,Ferrara:1996um,Ferrara:1997tw}, which governs the flow of the scalar fields as one approaches the black hole. Again, the possible attractor flows and consequently many key physical properties of the solutions, such as the  Bekestein-Hawking entropy and the degree of supersymmetry preserved,   can be understood using U-duality. See, for example, \cite{Ferrara:1997uz,Ferrara:2006xx,Andrianopoli:2006ub, Bellucci:2006xz, Borsten:2011ai} and the references therein. Clearly,  symmetries  are central to 
our understanding of black holes in supergavity.

In \cite{Borsten:2009zy} it was shown that when the U-duality group satisfies certain algebraic conditions,  in which case it is said to be of type $E_7$ \cite{Cartan, Brown:1969}, the black hole solutions enjoy a further discrete symmetry,  \emph{Freudenthal duality}. Such theories include all four-dimensional $\mathcal{N} > 2$ extended supergravities as well as   $\mathcal{N} = 2$ supergravity coupled to vector multiplets such that the scalars parametrise a symmetric space.  In these cases  the extremal static black holes carry electromagnetic charges transforming linearly in a symplectic representation of the U-duality group and the Bekenstein-Hawking entropy is a U-duality invariant function of these charges. For example, in $\mathcal{N}=8$ supergravity \cite{Cremmer:1979up} there are $28+28$ electric and magnetic charges transforming in the fundamental $\rep{56}$ of the U-duality group $E_{7(7)}$ and the Bekenstein-Hawking entropy is given by the unique quartic invariant of $E_{7(7)}$ \cite{Kallosh:1996uy}. The Freudenthal dual of a given set of black hole charges is a 
non-polynomial function of these charges, given in \eqref{def-2}, that  transforms in the same linear symplectic representation of the U-duality group, i.e. the $\rep{56}$ of $E_{7(7)}$ in the $\mathcal{N}=8$ case. Remarkably,  the Bekenstein-Hawking entropy is left invariant by the Freudenthal transformation despite its non-poloynomial character. Subsequently, F-duality was shown, through a suitable generalisation,  to be a symmetry not only of the  Bekenstein-Hawking entropy but also of the critical points of the black hole potential and was  extended to include any supergravity characterised by generalized special geometry \cite{Ferrara:2011gv}.

 Freudenthal duality has since been shown to play an important role in the structure of the extremal black hole solutions. For example,  the most general solution to the supersymmetric stabilization equations, entering in to the attractor mechanism,  in $
\mathcal{N}=8$, $D=4$ supergravity can be expressed in terms of the F-dual of a suitably defined real 56-dimensional vector $\mathcal{I}$, whose components are real harmonic functions in the Euclidean $\mathds{R}%
^{3}$ transverse space \cite{Ortin:2012gg}.

More recently, in \cite{Galli:2012ji}, Freudenthal duality  was also shown to be an on-shell
symmetry of the one-dimensional effective action that governs static,
spherically symmetric and asymptotically flat black hole solutions, both
extremal and non-extremal, in $\mathcal{N}=2$, $D=4$ supergravity. This  observation was used to establish the existence  of non-harmonic representations of the supersymmetric black hole solutions.

 In the present paper we prove that
the generalised scalar-dependent F-duality introduced in \cite%
{Ferrara:2011gv} is in fact a symmetry of the equations of motion of the full theory and is not
restricted to the extremal black hole solutions or their effective action. This
result holds for any ($\mathcal{N}\geqslant 2$, $D=4$) symplectic geometry
and thus goes beyond supergravities with U-duality group of type $E_{7}$.

The concept of F-duality is also applied here to world-sheet actions. The
Nambu-Goto world-sheet action in any $(t,s)$ signature spacetime can be
written in terms of the F-dual, making F-duality a manifest symmetry of
the action. The corresponding field equations and Bianchi identities are
then related by F-duality allowing for an F-dual formulation of
Gaillard-Zumino duality \cite{Gaillard:1981rj, Cecotti:1988zz, Duff:1989tf} on the world-sheet. Borrowing ideas from black hole
physics in supergravity an equivalent quadratic ``Polyakov-type'' action is
introduced using the black hole potential. Such a construction allows for
actions invariant under all groups of type $E_7$, including $E_7$ itself,
although in this case the stringy interpretation is less clear.

Before going into more details let us briefly review the key aspects of
F-duality as they stand in the current literature \cite{Borsten:2009zy,Ferrara:2011gv,Marrani:2012uu,Galli:2012ji}.

\subsection{Background: F-dual Black Holes}\label{sec:background}

The Freudenthal triple
system (FTS) was introduced in 1954 by Hans Freudenthal in order to understand
the  Lie group $E_{7}$, and its non-trivial minuscule representation, in terms of  the exceptional simple Jordan algebra \cite%
{Freudenthal:1954, Freudenthal:1959}. This structure is elegantly captured by the axiomatization  of the FTS \cite{Meyberg:1968,
Brown:1969}, described below, which may be regarded as
\emph{defining} a class of groups sharing  a common structure, as  encapsulated in the axioms. Such groups are said to be ``of type $E_{7}$''.  The essential properties of F-duality follow from the algebraic structure underlying the groups of type $
E_{7}$. From a more physical perspective these features can be traced
back to the generalised special geometry underpinning $\mathcal{N}\geq 2$
supergravity, which subsumes those theories of type $E_{7} $ \cite%
{Ferrara:2011gv}.

Axiomatically, a Freudenthal triple system is a vector space $\mathfrak{F}$, defined over a field\footnote{When needed we will denote an FTS defined over a specific field $\F$ by $\FTS_\F$.} $\mathds{F}$ not of characteristic 2 or 3,
equipped with three maps: (i) a non-degenerate bilinear antisymmetric form $%
\{x,y\}=-\{y,x\}\in \mathds{F}$, (ii) a not identically zero four-linear quartic
norm $\Delta (x,y,z,w)\in \mathds{F}$, (iii) a  trilinear
triple product $T(x,y,z)\in \mathfrak{F}$ defined\footnote{%
For convenience we have rescaled the usual definition of the quartic norm $%
q(x,y,z,w)=2\Delta (x,y,z,w)$, where $\{T(x,y,z),w\}=q(x,y,z,w)$.} by $%
\{T(x,y,z),w\}=2\Delta (x,y,z,w)$, where $x,y,z,w\in \mathfrak{F}$. These are then required to satisfy the  defining FTS relation,
\begin{equation}\label{eq:defFTS}
3\{T(x,x,y),T(y,y,y)\}=2\{x,y\}\Delta(x,y,y,y).
\end{equation}%
The \emph{automorphism} group $\Aut(\FTS)$, defined as the set of invertible $\F$-linear transformations preserving the quartic norm $\Delta$ and the symplectic form $\{x, y\}$, defines  a group of type $E_7$.

The original black hole F-dual  is defined on an \emph{integral} FTS  \cite{Borsten:2009zy, Krutelevich:2004}, an integral structure denoted $\FTS_\Z$ embedded in $\FTS_\R$, which breaks the automorphism group to a discrete subgroup $\Aut(%
\mathfrak{F}_\mathds{Z})$, e.g.\ $E_{7(7)}( \mathds{Z})$ in the case of $%
\mathcal{N}=8$ \cite{Hull:1994ys}.  It has two particularly interesting features: $(a)$ it is
an anti-involution $\tilde{\tilde{x}}=-x$, $(b)$ it leaves invariant the
leading-order Bekenstein-Hawking entropy. Moreover, in the maximally
supersymmetric $\mathcal{N}=8$ case there is evidence, deriving from an $%
E_{7(7)}(\mathds{Z})$-invariant dyon degeneracy formula valid on a subset of
1/8-BPS states \cite{Sen:2008sp}, that the entropy is in fact invariant
under \~{F}-duality to all orders \cite{Borsten:2009zy}.

Recently, groups of type $E_{7}$, Freudenthal triple systems, and F-duality  have also appeared in several indirectly related contexts. They have been investigated in
relation to minimal coupling of vectors and scalars in cosmology and
supergravity \cite{Ferrara:2011dz, Ferrara:2012qp}. F-duality also
recently appeared in \textit{Freudenthal gauge theory}\footnote{%
Recently, in \cite{Galli:2012ji} F-duality was proved to act as a
continuous gauge symmetry on the $H$-variables describing static
and extremal black holes of $\mathcal{N}=2$, $D=4$ supergravity. However, the resulting Freudenthal gauge symmetry is unrelated to the one constructed in  \cite{Marrani:2012uu}.}, where the scalar fields are $\FTS_\R$-valued, giving rise to gauge and global
symmetries  \cite{Marrani:2012uu}. Another application is in the context of entanglement in quantum information theory  \cite{Borsten:2008, levay-2008, Borsten:2009yb, LevayVrana, Borsten:2010ths, Szalay:2012},
with F-duality making a particularly important appearance in \cite{Levay:2012tg}. This is actually related to its application to black holes via the black-hole/qubit correspondence \cite{Borsten:2008wd, Borsten:2012fx}.

Classically ($\F=\R$), the
black hole charges $x$ are $\mathfrak{F}$-valued and transform linearly under
the automorphism group $\Aut(\mathfrak{F})$, which, modulo its two-element
centre, is isomorphic to the corresponding U-duality group. In this language
the black hole F-duality (which we will denote by \~{F}\footnote{Note, we  consider three variants of Freudenthal duality, which will be distinguished by three distinct notations for the F-dual: \~F, \^F and \u F.})
\begin{equation}
\text{\~{F}}:\mathfrak{F}_{0}\rightarrow \mathfrak{F}_{0},\qquad x\mapsto
\text{\~{F}}\left( x\right) =:\tilde{x},\qquad \text{where}\qquad \mathfrak{F%
}_{0}:=\{x\in \mathfrak{F}\;|\;\Delta(x,x,x,x)\neq 0\}  \label{def-1}
\end{equation}%
is defined by
\begin{equation}
\tilde{x}:=\sgn(\Delta (x))\frac{T(x)}{\sqrt{|\Delta (x)|}}=\nabla  \sqrt{|\Delta (x)|},  \label{def-2}
\end{equation}%
where $T(x)=T(x,x,x)$ and $\Delta (x)=\Delta (x,x,x,x)$ and  we have introduced the  gradient operator
\be
\nabla:=\Omega_{ab}\frac{\partial\phantom{x_b}}{\partial x_b} \label{def-grad}
\ee
in a  basis $\{e^a\}$ using the  symplectic metric  $\Omega$  defined by $\{x,y\}$ as in \eqref{eq:K}. The Bekenstein-Hawking entropy is given by
\begin{equation}
S_{\text{BH}}=\pi \sqrt{|\Delta (x)|}=\pi \sqrt{|\Delta (\tilde{x})|}
\end{equation}%

Quantum
mechanically the charges $x$ are subject to the Dirac-Schwinger quantisation
conditions and so lie on a lattice. Consequently, as previously emphasised, they must be
assigned to elements of an \emph{integral} FTS $\mathfrak{F}_\mathds{Z}=%
\mathfrak{F}(\mathfrak{J}_\mathds{Z})$ where $\mathfrak{J}_\mathds{Z}$ is an
\emph{integral} cubic Jordan algebra \cite%
{Borsten:2009zy, Elkies:1996,Gross:1996,Krutelevich:2002, Krutelevich:2004}. The
corresponding U-duality is given by the discrete automorphism group $\Aut(%
\mathfrak{F}_\mathds{Z})$, defined as the set of invertible $\Z$-linear transformations preserving the quartic norm $\Delta$ and the symplectic form $\{x, y\}$, which are $\Z$-valued in this case.
This integral structure introduces several subtleties, in particular, not every charge configuration has a well defined \~F-dual  \cite{Borsten:2009zy}. Moreover, in the classical ($\FTS_\R$) case every \~F-dual pair of charge vectors is also U-dual, whereas in the quantum ($\FTS_\Z$) case this is no longer true. However, these consideration will be largely overlooked here, suffice to say that our results  are equally valid in either case, $\Aut(\FTS_\R)$ vs. $\Aut(\FTS_\Z)$. The interested reader can refer to \cite{Borsten:2009zy, Bianchi:2009mj, Borsten:2010aa} for   more on the discrete story.

We will refer to  the \~{F}-dual as  ``critical'' in
the sense that it is defined in terms of only the charges and is independent of the scalar
fields. In \cite{Ferrara:2011gv} an interpolating
scalar-dependent  ``non-critical'' formulation
of the F-duality (which we will denote by \^{F})
\begin{equation}
\text{\^{F}}:\mathfrak{F}_{0}\rightarrow \mathfrak{F}_{0},\qquad x\mapsto
\text{\^{F}}\left( x\right) =:\hat{x},  \label{F-hat-1}
\end{equation}%
was introduced using the so-called effective black hole potential \cite%
{Gibbons:1996af} $V_{\text{BH}}(\phi ,x)$ :
\begin{equation}
\hat{x}(\phi ):=\nabla V_{\text{BH}}(\phi ,x),\qquad \text{where}%
\qquad V_{\text{BH}}(\phi ,x):=-\frac{1}{2}\{x,\mathcal{S}(\phi )x\}.
\label{F-hat-2}
\end{equation}%
in the conventional FTS (but unconventional supergravity\footnote{%
Our conventions are explained in detail in \autoref{sec:fb}.}) notation.

For U-duality groups of type $E_7$ $\mathcal{S}(\phi )=\Omega \mathcal{M}(\phi)\in \Aut(\mathfrak{F})$ is
a scalar-dependent almost complex structure which may be regarded as the
projection onto the adjoint in the symmetric tensor product of the
representation carried by $\mathfrak{F}$. However,  since the \^F-dual is defined in terms of the black hole potential only, it is consistent for any ($\mathcal{N}\geqslant 2$, $D=4$) symplectic geometry and thus goes beyond supergravities with U-duality group of type $E_{7}$ \cite{Ferrara:2011gv}, as will be explained in \autoref{Beyond}. The generalised {\^{F}}-dual is
again an anti-involution $\hat{\hat{x}}=-x$ and, moreover, coincides with the
scalar-independent definition when evaluated at the black hole horizon, i.e. at the critical points of the black hole potential,
\begin{equation}
\hat{x}_{\ast }=\hat{x}(\phi _{\ast })=\tilde{x},
\end{equation}%
where $\phi _{\ast }$ denotes the horizon value of the moduli. Note, the
black hole potential is itself invariant under the generalised \^{F}-dual
\cite{Ferrara:2011gv}
\begin{equation}
V_{\text{BH}}(\phi ,\hat{x})=V_{\text{BH}}(\phi ,x),  \label{V-inv-2}
\end{equation}%
which follows, in the case of groups of type $E_7$, from $\{\sigma x,\sigma
y\}=\{x,y\},\forall \sigma \in \Aut(\mathfrak{F})$.  This
implies that while two \^{F}-dual black holes have generically distinct
off-shell effective potentials $V(\phi ,x)\not=V(\phi ,\tilde{x})$, their
entropy is the same despite the fact their (attractor) scalar flows are
typically different.

\subsection{Beyond Black Hole Solutions}

Thus far our discussion of F-duality has been confined to a rather
restricted class of black hole solutions, initially focussing on the charges
carried by those solutions and then in terms of the effective potential
describing their scalar dynamics. This raises the natural question, which we aim to address in the present work:  can
F-duality be reformulated as a symmetry of the full theory or is
it only defined at the level of the extremal black hole solutions?

In response we will show in \autoref{sec:bh3} that, for those theories with
U-dualities of type $E_{7}$, the scalar-dependent \^{F}-dual can be promoted to a symmetry of the theory itself. Since Freudenthal duality relies crucially on the  properties of the
representation carried by $\mathfrak{F}$, formulations of the  supergravity  (specifically its bosonic sector)
which make  U-duality manifest are the most convenient set-ups with which to address this question. The doubled Lagrangian framework of
\cite{Cremmer:1997ct}, in which the Lagrangian is written in terms of
doubled gauge potentials treated as independent fields but supplemented by a
U-duality invariant constraint equation, provides such a construction. We
emphasise that the price one must pay in order to write down a
simultaneously Lorentz and U-duality invariant Lagrangian is the need for a
constraint, which cannot be derived from the Lagrangian and so must be
imposed by hand. As such it is essentially defined only at the classical level and its quantisation remains unclear. For our purposes, however, the
doubled Lagrangian formalism simply provides a convenient framework to make
F-invariance manifest. Symmetries of the doubled Lagrangian and its constraint are classical   symmetries of the corresponding supergravity. Defining the \^{F}-duality operation on the $%
\mathfrak{F}$-valued doubled field strengths, it is easily seen to be a
symmetry of the doubled Lagrangian and of the constraint itself.
Hence, we may conclude that F-duality is  a symmetry of the supergravity
equations of motion and it is \emph{not} restricted to their black hole
solutions.

F-duality is actually a symmetry of another class of theories: Nambu-Goto world-sheet actions in a spacetime with signature $(t,s)$.  In \cite{Duff:2006ev} it was shown that the Nambu-Goto action in $(2,2)$ signature spacetime could be written as the square root of a quartic norm   given by Cayley's hyperdeterminant  \cite{Cayley:1845}, which belongs to an FTS. In this context it was used to illustrate a hidden triality invariance that is specific to $(2,2)$  signature.  However, Cayley's hyperdeterminant  is the quartic norm of but one FTS in a countably infinite sequence with automorphism group $\SL(2, \R)\times\SO(t,s)$. These  are analogously  associated with Nambu-Goto world-sheet actions in $(t,s)$ signature, as is made explicit in \autoref{sec:ws}. Here, the $\SL(2, \R)$ factor corresponds to a global subgroup of the world-sheet diffeomorphisms. The Nambu-Goto actions can then be written in terms of the world-sheet 1-forms $F$ and their Freudenthal duals $\tilde{F}$,
\begin{equation}
S_{\text{NG}}=\frac{1}{2}\int d^{2}\xi\{\tilde{F},F\}\end{equation}%
which makes \~F-duality manifest. Introducing a set of auxiliary scalar fields, parametrising $[\SL(2,\R)\times\SO(t,s)]/[\SO(2)\times \SO(t,s)]$, and using the idea of the black hole potential we can also define the ``non-critical'' \^F-dual, $F\mapsto \hat{F}$.

Having described the Nambu-Goto action in $(t,s)$ signature in terms of a specific  class of FTS, it is natural ask whether  other FTS could be used in this stringy context. Indeed, recalling that Cartan's quartic $E_{7(7)}$ invariant $I_4$ reduces to  Cayley's hyperdeterminant in a canonical basis \cite{Kallosh:2006zs},  it was already suggested in \cite{Duff:2006ev} that $I_4$  provides a natural generalisation of the Nambu-Goto action in $(2,2)$ signature with an $\SO(6,6)$ Lorentz group  embedded in the larger $E_{7(7)}$ symmetry. Here, this idea is developed  in \autoref{sec:E7string}. In principle one could consider any FTS, however, we focus on the ``maximal'' case of $\FOs$, where    $F\in \FOs$ transforms as the  fundamental $\rep{56}$ of $\Aut(\FOs)=E_{7(7)}$ and $\Delta(F)=I_4(F)$. The notation $\FOs$ refers to the fact that it may be constructed using the Jordan algebra of $3\times 3$ Hermitian matrices defined over the split octonions $\Oct^s$. This gives a manifestly $E_{7(7)}$-invariant ``world-sheet'' action. However, in order to clarify its stringy interpretation, we decompose the $\rep{56}$ w.r.t. $\SL(2,\R)\times\SO(6,6)$, yielding a  world-sheet 1-form $F_{\alpha}^{a}$ in a target space with $(6,6)$ signature coupled to 32 auxiliary world-sheet scalar densities  (a target space Weyl spinor) $\lambda^A$.  These two fields are mixed by the larger global $E_{7(7)}$ symmetry.

\subsection{Plan}

In \autoref{sec:fb} we review the key properties of Freudenthal triple
systems and the three known incarnations of Freudenthal duality (i) the
``critical'' \~F-duality, which is defined on the black hole charges of
supergravities with U-duality groups type $E_7$, (ii) the ``critical'' \u{F}%
-duality, which is defined on the black hole charges of all $\mathcal{N}%
\geqslant 2$-extended supergravity theories in $D=4$ using the entropy
function, (iii) the scalar-dependent ``non-critical'' \^{F}-duality, which is
defined on the black hole charges of all $\mathcal{N}\geqslant 2$-extended
supergravity theories in $D=4$ using the black hole potential away from its
critical points. Furthermore, in \autoref{sec:bh3} we recall the definition of the doubled Lagrangian
formalism, and use it to show \^{F}-duality is an invariance of the equations
of motion.

In \autoref{sec:ws} we show that the Nambu-Goto action naturally defines an
FTS and that it can therefore be written in terms of the \~F-dual. This
yields an \~F-dual formulation of Gaillard-Zumino duality on the
world-sheet. We conclude with some more speculative observations on the
existence of world-sheet actions with $\Aut(\mathfrak{F})$-symmetry for
generic $\mathfrak{F}$, including the exceptional case of $E_7$.

\section{Freudenthal Duality}

\label{sec:fb}

\subsection{The Freudenthal Triple System}

In 1954 Freudenthal \cite{Freudenthal:1954, Freudenthal:1959} found that the
133-dimensional exceptional Lie group $E_{7}$ could be understood in terms
of the automorphisms of a construction based on the minimal dimensional $%
E_{7}$-module $\mathbf{56}$ built from the exceptional Jordan algebra of $%
3\times 3$ Hermitian octonionic matrices. Today this construction goes by
the name of \emph{the Freudenthal triple system}, reflecting the special
role played by its triple product.

Following Freudenthal, Meyberg \cite{Meyberg:1968} and Brown \cite%
{Brown:1969} axiomatized the ternary structure underlying the FTS. The $E_7$%
-module is just one of a class of modules of ``groups of type $E_7$''. The
FTS carries the representation of the dyonic black hole charge vectors for a
broad class of 4-dimensional supergravity theories \cite{Gunaydin:1983bi,
Gunaydin:1983rk, Ferrara:1997uz}.

An FTS is axiomatically defined \cite{Brown:1969} as a finite dimensional
vector space $\mathfrak{F}$ over a field $\mathds{F}$ (not of characteristic
2 or 3), such that:

\begin{enumerate}
\item $\mathfrak{F}$ possesses a non-degenerate antisymmetric bilinear form $%
\{x, y\}.$

\item $\mathfrak{F}$ possesses a symmetric four-linear form $\Delta(x,y,z,w)$
which is not identically zero.

\item If the ternary product $T(x,y,z)$ is defined on $\mathfrak{F}$ by $%
\{T(x,y,z), w\}=2\Delta(x, y, z, w)$, then
\begin{equation}  \label{eq:def}
3\{T(x, x, y), T(y,y,y)\}=2\{x, y\}\Delta(x, y, y, y).
\end{equation}
\end{enumerate}

It will often be notationally convenient to introduce a basis $\{e^{a}\}$, $%
a=1,\ldots \dim _{\mathds{F}}\mathfrak{F}$ such that
\begin{equation}\label{eq:K}
\{e^{a},e^{b}\}=\Omega ^{ab},\qquad \Delta(e^{a},e^{b},e^{c},e^{d})=K^{abcd}.
\end{equation}%
Since $\{x ,y \}$ is non-degenerate, we can define the
symplectic 2-form $\Omega _{ab}$ such that $\Omega _{ab}\Omega ^{bc}=\delta
_{a}{}^{c}$. We will raise/lower indices with $\Omega ^{ab}/\Omega _{ab}$.

The \emph{automorphism} group of an FTS is defined as the set of invertible $%
\mathds{F}$-linear transformations preserving the quartic and quadratic
forms:
\begin{equation}
\Aut(\mathfrak{F}):=\{\sigma \in \Iso_{\mathds{F}}(\mathfrak{F})|\{\sigma
x,\sigma y\}=\{x,y\},\;q(\sigma x)=q(x)\}.  \label{eq:brownfts}
\end{equation}%
Note, the conditions $\{\sigma x,\sigma y\}=\{x,y\}$ and $q(\sigma x)=q(x)$
immediately imply
\begin{equation}
T(\sigma x)=\sigma T(x).
\end{equation}

A remarkable result due to Brown \cite{Brown:1969} and Ferrar \cite%
{Ferrar:1972} implies that every simple reduced\footnote{%
An FTS is simple if and only if $\{x ,y \}$ is non-degenerate,
which we assume. An FTS is said to be reduced if it contains a strictly
regular element: $\exists \;u\in \mathfrak{F}$ such that $T(u,u,u)=0$ and $%
u\in \text{ Range }L_{u,u}$ where $L_{x,y}:\mathfrak{F}\rightarrow \mathfrak{%
F};\quad L_{x,y}(z):=T(x,y,z)$. Note that FTS on \textquotedblleft
degenerate" groups of type $E_{7}$ (as defined in \cite{Ferrara:2012qp}, and
Refs. therein) are not reduced and hence cannot be written as $\mathfrak{F}(%
\mathfrak{J})$; they correspond to theories which cannot be uplifted to $D=5$
dimensions consistently reflecting the lack of an underlying $\mathfrak{J}$.}
FTS $\mathfrak{F}$ is isomorphic to an FTS $\mathfrak{F}(\mathfrak{J})$,
where
\begin{equation}
\mathfrak{F}(\mathfrak{J}):=\mathds{F}\oplus \mathds{F}\oplus \mathfrak{J}%
\oplus \mathfrak{J}
\end{equation}%
and $\mathfrak{J}$ is the Jordan algebra of an admissible cubic form with
base point or the Jordan algebra of a non-degenerate quadratic form. The FTS
quadratic form, quartic norm and triple product are then defined in terms of
the basic Jordan algebra operations \cite{Brown:1969, Ferrar:1972}; for
details, see \cite{Borsten:2011nq} and Refs. therein. We will not make
explicit use of this perspective here, however, it is useful for tabulating
the automorphism groups according as the underlying cubic Jordan algebra as
we have done in \autoref{tab:FTSsummary}. From a physical point of view, the
underlying Jordan algebra structure originates from the fact that the $D=4$
supergravities may be obtained by dimensionally reducing a $D=5$ theory
whose U-duality is characterised by the reduced structure group of the
corresponding Jordan algebra.

For $\mathfrak{F}(\mathfrak{J}_{3}^{\mathds{A}})$ the automorphism group has
a two element centre and its quotient yields the simple groups listed in %
\autoref{tab:FTSsummary}, while for $\mathfrak{F}(\mathfrak{J}_{m,n})$ one
obtains the semi-simple groups $\SL(2,\mathds{R})\times \SO(m+1,n+1)$ \cite%
{Brown:1969,Krutelevich:2004, Gunaydin:2009zza}. In all cases $\mathfrak{F}$
forms a symplectic representation of $\Aut{(\FTS)}$, the dimensions of which
are listed in the final column of \autoref{tab:FTSsummary}. This table
covers a number 4-dimensional supergravities: $\mathfrak{F}^{2,n}:=\mathfrak{%
F}\left( \mathfrak{J}_{1,n-1}\right) $ and $\mathfrak{F}^{6,n}:=\mathfrak{F}%
\left( \mathfrak{J}_{5,n-1}\right) $ respectively correspond to $\mathcal{N}%
=2,4$ Maxwell-Einstein supergravity, while $\mathfrak{F}^{\mathds{A}}:=%
\mathfrak{F}\left( \mathfrak{J}_{3}^{\mathds{A}}\right) $ correspond to $%
\mathcal{N}=2$ \textquotedblleft magic\textquotedblright\ Maxwell-Einstein
supergravity, $\mathfrak{F}^{\mathds{O}^{s}}:=\mathfrak{F}\left( \mathfrak{J}%
_{3}^{\mathds{O}^{s}}\right) $ corresponds to $\mathcal{N}=8$ maximally
supersymmetric supergravity, and $\mathfrak{F}\left( \mathds{R}\right) $
corresponds to the $\mathcal{N}=2$ $t^{3}$ model (see, for example, \cite%
{Gunaydin:1983bi,Gunaydin:1983rk,Rios:2007qn,Borsten:2008wd,Borsten:2009zy,Borsten:2010aa,Rios:2010br,ICL-2,Rios:2011fa}%
).

\begin{table}[tbp]
\caption[Jordan algebras, corresponding FTSs, and their associated symmetry
groups]{The automorphism group $\Aut(\mathfrak{F}(\mathfrak{J}))$ and the
dimension of its representation $\dim\mathfrak{F}(\mathfrak{J})$ given by
the Freudenthal construction defined over the cubic Jordan algebra $%
\mathfrak{J}$ over $\mathds{R}$ (with dimension $\dim\mathfrak{J}$ and
reduced structure group $\Str_0(\mathfrak{J})$).}
\label{tab:FTSsummary}%
\begin{tabular*}{\textwidth}{@{\extracolsep{\fill}}c*{5}{M{c}}c}
\toprule
& \text{Jordan algebra }\mathfrak{J} & \Str_0(\mathfrak{J}) & \dim\mathfrak{J} & \Aut(\mathfrak{F}(\mathfrak{J})) & \dim\mathfrak{F}(\mathfrak{J}) &\\
\hline
& \mathds{R}                     & -                                  & 1   & \SL(2,\mathds{R})                                   & 4    & \\
& \mathds{R}\oplus\mathds{R}           & \SO(1,1)                       & 2   & \SL(2,\mathds{R})\times \SL(2,\mathds{R})                  & 6    & \\
& \mathds{R}\oplus\mathds{R}\oplus\mathds{R} & \SO(1,1)\times \SO(1,1)    & 3   & \SL(2,\mathds{R})\times \SL(2,\mathds{R})\times \SL(2,\mathds{R}) & 8    & \\
& \mathds{R}\oplus \Gamma_{1,n-1}           & \SO(1,1)\times \SO(1,n-1)  & n+1 & \SL(2,\mathds{R})\times \SO(2,n)                & 2(n+2) & \\
& \mathds{R}\oplus \Gamma_{5,n-1}           & \SO(1,1)\times \SO(5,n-1)  & n+5 & \SL(2,\mathds{R})\times \SO(6,n)                & 2(n+6) & \\
& \mathfrak{J}_{3}^{\R}             & \SL(3, \mathds{R})                         & 6   & \Sp(6,\mathds{R})                                   & 14   & \\
& \mathfrak{J}_{3}^{\C}             & \SL(3,\mathds{C})                         & 9   & \SU(3,3)                                 & 20   & \\
& \mathfrak{J}_{3}^{\C^s}             & \SL(3,\mathds{R})\times\SL(3, \mathds{R})                         & 9   & \SL(6, \mathds{R})                                 & 20   & \\
& \mathfrak{J}_{3}^{\Q}             & \SU^\star(6)                   & 15  & \SO^\star(12)                            & 32   & \\
& \mathfrak{J}_{3}^{\Q^s}             & \SL(6, \mathds{R})                   & 15  & \SO(6,6)                            & 32   & \\
& \mathfrak{J}_{3}^{\Oct}             & E_{6(-26)}                   & 27  & E_{7(-25)}                            & 56   & \\
& \mathfrak{J}_{3}^{\Oct^s}             & E_{6(6)}                   & 27  & E_{7(7)}                            & 56   & \\
\bottomrule
\end{tabular*}
\end{table}

\subsection{\label{F-tilde} Black Hole \~{F}-Duality...}

The black hole \~F-duality is a non-polynomial
transformation on the electromagnetic charges $x$ of extremal single-centre
large black hole solutions in $D=4$ supergravities with U-duality group of
type $E_{7}$ \cite{Borsten:2009zy}.

The essential
characteristics of the \~F-dual,  defined by \eqref{def-1} and \eqref{def-2}, follow from the defining FTS relation
\eqref{eq:defFTS},
which, in particular, implies \cite{Brown:1969}
\begin{equation}\label{eq:TofTofx}
T(T(x))=-\Delta ^{2}(x)x.
\end{equation}%

The invariance of $\Delta (x)$ follows by recalling
that
\begin{equation}
2\Delta (x)=\{T(x),x\}.
\end{equation}%
 Hence, using \eqref{eq:TofTofx}
\begin{equation}
\Delta (T(x))=\Delta (x)^{3}
\end{equation}%
one finds
\begin{equation}
\Delta (\tilde{x})=\Delta (T(x))\Delta (x)^{-2}=\Delta (x).
\end{equation}%
Moreover, \~{F}-dual is anti-involutive,
\begin{equation}
\tilde{\tilde{x}}=T(\tilde{x})|\Delta (x)|^{-1/2}=T(T(x))\Delta (x)^{-2}=-x.
\end{equation}
Note in particular, that from the above we have%
\begin{equation}
\{\tilde{x},x\}=2\sqrt{\left\vert \Delta (x)\right\vert }.
\label{sympl-x-x-tilde}
\end{equation}
Since the Bekenstein-Hawking entropy is given by,
\begin{equation}
S_{\text{BH}}=\pi \sqrt{|\Delta (x)|}=\frac{\pi}{2}\{\tilde{x},x\},
\end{equation}
it is manifestly invariant under F-duality.

For classical black hole solutions the charges are $\mathds{R}$-valued,
hence    $\mathfrak{F}$ is defined over $\mathds{R}$. However, quantum
mechanically this is no longer the case. The Dirac-Schwinger quantisation
condition relating two black holes with charges $x$ and $x^{\prime }$ is
given by
\begin{equation}  \label{eq:DS}
\{x, x^{\prime }\}\in \mathds{Z}.
\end{equation}
Consequently, the black hole charges $x$ must be
assigned to elements of an \emph{integral} FTS $\mathfrak{F}_\mathds{Z}=%
\mathfrak{F}(\mathfrak{J}_\mathds{Z})$ where $\mathfrak{J}_\mathds{Z}$ is an
\emph{integral} cubic Jordan algebra \cite%
{Elkies:1996,Gross:1996,Krutelevich:2002, Krutelevich:2004}. The
corresponding U-duality is given by the discrete automorphism group $\Aut(%
\mathfrak{F}_\mathds{Z})$, e.g.\ $E_{7(7)}( \mathds{Z})$ in the case of $%
\mathcal{N}=8$ \cite{Hull:1994ys}. In particular, $\Delta(x)$ is now
quantised:
\begin{equation}
\Delta(x) \in \{0,1\} \mod 4.
\end{equation}

This integral structure constrains the class of black holes admitting a
well-defined \~F-dual. Requiring that $\tilde x$ be integral restricts us to
that subset of black holes for which $|\Delta(x)|$ is a perfect square and
for which $|\Delta(x)|^{1/2}$ divides $T(x)$:
\begin{equation}  \label{eq:d4}
d_4(x)=\left [\frac{d_3(x)}{d_1(\tilde x)}\right]^2,
\end{equation}
where we have introduced a set of discrete U-duality invariants constructed
in terms of the greatest common divisor (gcd) \cite%
{Krutelevich:2004,Sen:2008sp,Borsten:2009zy}:
\begin{equation}  \label{eq:DiscreteInvariants}
\begin{split}
d_1(x)&=\gcd(x) \\
d_2(x)&=\gcd(3\,T(x,x,y)+\{x,y\}\,x) \ \forall\ y \\
d_3(x)&=\gcd( T(x,x,x) ) \\
d_4(x)&=|\Delta(x)| \\
d^{\prime }_4(x)&=\gcd(x \wedge T(x)).
\end{split}%
\end{equation}

In the classical theory, where the black hole charges are real-valued, any
pair of \~F-dual charge vectors are related by U-duality. However, in the
quantum theory it is no longer true that any well defined pair of \~F-dual
black holes are U-dual \cite{Borsten:2009zy}. This is most directly
confirmed by considering the discrete U-duality invariants listed in %
\eqref{eq:DiscreteInvariants}. While $d_2(x)$, $d_4(x)$, and $d^{\prime
}_4(x)$ are \~F-dual invariant $d_1=\gcd(x)$ and $d_3=\gcd(T(x))$ need not
be. Since corrections to the leading-order Bekenstein-Hawking entropy may
depend on the discrete invariants \cite{Maldacena:1999bp,
Dabholkar:2007vk,Sen:2007qy,Banerjee:2007sr,Banerjee:2008pu,Banerjee:2008ri,Sen:2008ta,Sen:2008sp,Bianchi:2009mj}
it is not clear that \~F-duality is a symmetry to all orders. However, for
a   subclass, closed under $E_{7(7)}(\mathds{Z})$ and F-duality, of 1/8-BPS dyons in type II
string theory on a 6-torus Sen \cite{Sen:2008sp} derived degeneracy formula
which is a function of only $d_4$ and $d^{\prime }_4$ and therefore \emph{is}
\~F-dual invariant.

\subsection{\label{Beyond}...and its Generalization Beyond Groups of Type $%
E_{7}$: \u{F}-Duality}

The symplectic structure of $\mathcal{N}\geqslant 2$-extended supergravity
theories in $D=4$ spacetime dimensions \cite{Ceresole:1995ca,
Ferrara:2006em} allows one to generalize the ``critical''  \~{F}%
-duality \textit{beyond} U-duality groups of type $E_{7}$; we will here
denote such a generalization \u{F}-duality. Its action on $Sp\left( 2n,\mathbb{R}\right) $ charge vectors $x$ is defined
as \cite{Ferrara:2011gv} (recall definition (\ref{def-grad}))%
\begin{equation}
\breve{F}:x\longmapsto \breve{F}\left( x\right) =:\breve{x}:=\nabla S\left(
x\right) =-\Omega \mathcal{M}\left( \phi _{\ast }\left( x\right) \right) x.%
\label{14}
\end{equation}%
By indicating with $\phi $ the set of scalar fields, $\mathcal{M}$ denotes
the scalar-dependent, symplectic, negative definite, real symmetric matrix
made of the vector couplings \cite{Breitenlohner:1987dg, Ceresole:1995ca}:%
\begin{equation}
\mathcal{M}( \phi ) \Omega \mathcal{M}( \phi ) =\Omega
,~\mathcal{M}^{T}( \phi ) ~=\mathcal{M}( \phi ) .
\label{M-sympl}
\end{equation}%
Furthermore, $S(x) $ denotes the Bekenstein-Hawking black hole
entropy (homogeneous of degree two in the charges $x$):%
\begin{equation}
S( x) =\pi V_{\text{BH}}( \phi _{\ast }( x)
,x) ,  \label{SBH}
\end{equation}%
where $\phi _{\ast }( x) $ denotes the $x$-dependent critical
values of the scalar fields, defined as
\begin{equation}
\left. \partial _{\phi }V_{\text{BH}}\left( \phi ,x\right) \right\vert
_{\phi _{\ast }\left( x\right) }=0.  \label{V-crit}
\end{equation}%
It can be proved \cite{Ferrara:2011gv} that the critical points $\phi _{\ast
}( x) $ of $V_{\text{BH}}( \phi ,x) $ (defined by (\ref%
{V-crit})) coincide with the critical points $\phi _{\ast }( \breve{x} ) $ of \u{F}$( V_{\text{BH}}( \phi
,x) ) :=V_{\text{BH}}( \phi ,\breve{x}
) $ (defined by (\ref{V-crit}) with $x\rightarrow \breve{x}$). It this follows that the symplecticity of $\mathcal{M}$ (\ref%
{M-sympl}) (in particular, evaluated at $\phi =\phi _{\ast }\left( x\right)
=\phi _{\ast }(\breve{x}) $), implies the
anti-involutivity of \u{F}-duality (\ref{14}):%
\begin{equation}
\text{\u{F}}^{2}:x\longmapsto \text{\u{F}}^{2}\left( x\right) :=\text{\u{F}}%
\left( \text{\u{F}}\left( x\right) \right) =-\Omega \mathcal{M}\left( \phi
_{\ast }(\breve{x}\right)) \breve{x}
=\Omega \mathcal{M}\left( \phi _{\ast }(x)\right) \Omega \mathcal{M}\left(
\phi _{\ast }(x)\right) x=-x.
\end{equation}

An important consequence of these results is that the Bekenstein-Hawking
entropy $S\left( x\right) $, which generally is a complicated non-polynomial
function homogenous of degree two in the charges $x$, is \u{F}-invariant
\cite{Ferrara:2011gv}:%
\begin{equation}
S( x) =\pi V_{\text{BH}}( \phi _{\ast }( x)
,x) =\pi V_{\text{BH}}( \phi _{\ast }( \breve{x} ) ,\breve{x}) =S(\breve{x}) \Leftrightarrow S(x)=S\left( \nabla S\left( x\right) \right) .  \label{S-inv}
\end{equation}%
(\ref{S-inv}) is a general result, which holds in any ($\mathcal{N}\geqslant
2$, $d=4$) generalized special geometry \cite{Ferrara:2006em, Ferrara:2011gv}%
. Within this broad class of theories, two \u{F}-dual black holes (namely,
two black holes whose dyonic charge vectors $x$ are related by \u{F}-duality
(\ref{14})) have generically different  effective potentials $V_{\text{BH}%
}( \phi ,x) \neq V_{\text{BH}}( \phi ,\breve{x} ) $, which however exhibit the same critical points ($\phi
_{\ast }\left( x\right) =\phi _{\ast }\left( \breve{x}
\right) $); thus, as yielded by (\ref{S-inv}), two \u{F}-dual black holes
have the same Bekenstein-Hawking entropy, despite the fact their (attractor)
scalar flows are generally different.

For those generalized special geometries \cite{Ferrara:2006em} related to
groups of type $E_{7}$ \cite{Brown:1969}, \u{F}-duality (\ref{14})
consistently reduces to the non-polynomial (homogeneity-preserving) \~{F}-duality defined by (\ref{def-1})-(\ref%
{def-2}). In particular, $S\left( x\right) $ reduces to the ($\pi $ times)
the square root of the (absolute value of) the quartic polynomial $\Delta
\left( x\right) $:%
\begin{equation}
\text{groups~of~type~}E_{7}:\text{~\u{F}}=\text{\~{F}}\Rightarrow S(x)=\pi
\sqrt{\left\vert \Delta \left( x\right) \right\vert }.
\end{equation}

\subsection{Generalised Scalar-Dependent Freudenthal Duality \^{F}}

\label{sec:bh2}

The scalar-dependent, \textquotedblleft non-critical \textquotedblright\
F-duality (denoted by \^{F}) was introduced in \cite{Ferrara:2011gv}, and it
is defined by (\ref{F-hat-1}) and (\ref{F-hat-2}), with
\begin{equation}
\text{\^{F}}\left( x\right) =:\hat{x}(\phi ):=\partial _{x}V_{\text{BH}%
}(\phi ,x)=-\mathcal{S}(\phi )x,\qquad \text{where}\qquad \mathcal{S}(\phi
):=\Omega \mathcal{M}\left( \phi \right) .  \label{F-hat-3}
\end{equation}%
Note that the scalar fields $\phi $ do not transform under \^{F}-duality.
From its very definition, the real matrix $\mathcal{S}(\phi )\in \Aut(%
\mathfrak{F})$ is a scalar-dependent almost complex structure which may be
regarded as the projection onto the adjoint in the symmetric tensor product
of the representation carried by $\mathfrak{F}$:%
\begin{equation}
\mathcal{S}^{2}=-\mathbb{I}.
\end{equation}

As a consequence of (\ref{M-sympl}), it is immediate to see that \^{F} is
anti-involutive:
\begin{equation}
\widehat{\widehat{x}}(\phi )=\Omega \mathcal{M}\left( \phi \right) \Omega
\mathcal{M}\left( \phi \right) x=-x.  \label{31}
\end{equation}%
Furthermore, \^{F}-duality coincides with \u{F}-duality when evaluated at
the black hole horizon
\begin{equation}
\hat{x}_{\ast }=\hat{x}(\phi _{\ast })=\text{\u{F}}\left( x\right) ,
\end{equation}%
where $\phi _{\ast }$ denotes the horizon value of the scalar fields,
defined by (\ref{V-crit}).

As given by (\ref{V-inv-2}) \cite{Ferrara:2011gv}, the black hole potential $%
V_{\text{BH}}$ is itself invariant under the generalised \^{F}-dual; this
result is general: it holds in any ($\mathcal{N}\geqslant 2$, $d=4$)
symplectic geometry \cite{Ferrara:2006em}.\textit{\smallskip }

It can also be checked that all definitions and results concerning \^{F}%
-duality consistently reduce, when evaluated at the (non-degenerate)
critical points of $V_{\text{BH}}$, to the analogous definitions and results
for \u{F}-duality, introduced in \autoref{Beyond}.

\subsection{\label{sec:bh3} \^{F}-Duality and Doubled Formalism in Supergravity}


The doubled Lagrangian
\begin{equation}
\mathcal{L}_{\text{double}}=R\star 1+\frac{1}{4}\tr(d\mathcal{M}^{-1}\wedge
\star d\mathcal{M})-\frac{1}{4}H\wedge \mathcal{M}\star {H}
\label{eq:genericDL}
\end{equation}%
and the twisted self-duality constraint
\begin{equation}
{H}=\Omega \mathcal{M}\star {H\label{constr}}
\end{equation}%
are equivalent at the level of equations of motion to the standard
formulation of supergravity when scalars parametrise a homogeneous symmetric space $G_4/H_4$ \cite{Cremmer:1997ct}. For the vector equations
of motion, the Lagrangian $\mathcal{L}_{\text{double}}$ should be varied
with respect to the doubled gauge potentials $(A,B)$, where $H=d(A,B)$,
treated as \textit{independent} fields; the constraint (\ref{constr}) is
then applied to the equations of motion. Similarly, for the scalar equations
of motion one first varies treating $H=(dA,dB)$ as independent and then
applies the constraint (\ref{constr}).

The \textquotedblleft non-critical \textquotedblright\  \^{F}-duality introduced in  \autoref{sec:bh2} may be extended to the doubled field strengths as follows (\textit{%
cfr.} (\ref{F-hat-3})):
\begin{equation}
\hat{H}:=-\mathcal{S}(\phi )H,
\end{equation}%
such that the constraint (\ref{constr}) can be rewritten as%
\begin{equation}
H=-\star \hat{H}.
\end{equation}
Since $\mathcal{M}=-\Omega \mathcal{S}$ and $\mathcal{S}^{2}=-\mathds{1}$,
it is clear that \^{F}-duality defined on the doubled field strengths leaves
$\mathcal{L}_{\text{double}}$ invariant (note that \^{F}-duality is inert on
the Einstein-Hilbert and non-linear sigma model terms). It is also clear
that
\begin{equation}
\hat{H}=\Omega \mathcal{M}\star \hat{H}\quad \Leftrightarrow \quad {H}%
=\Omega \mathcal{M}\star {H,}
\end{equation}%
and hence \^{F}-duality is a symmetry of the full theory, not just of the
black hole solutions.

It is here worth mentioning that one may consider to replace the last
term $-\frac{1}{4}H\wedge \mathcal{M}\star {H}$ of Eq.
(\ref{eq:genericDL}) with a purely H-dependent term $\sqrt{\left\vert
\Delta \left( H\right)
\right\vert }$. In this latter term, the structure of the contractions of
D=4 space-time indices of doubled
field strengths 2-forms $H$ is given by the
rank-$8$ tensor $t^{(8)}$ \cite{Schwarz:1982jn} (see also
\textit{e.g.} \cite{Chemissany:2011yv,Kallosh:2012yy}). It would be
interesting to study the possibility to formulate gravity theories
(beyond supergravity) within the doubled Lagrangian formalism by
exploiting the symmetries of $\Delta \left( H\right) $.

\section{Freudenthal-Dual World-Sheet Actions}

\label{sec:ws}

\subsection{The Nambu-Goto Action}

We start by considering the Nambu-Goto world-sheet action in a
pseudo-Euclidean spacetime with signature $(t,s)$,
\begin{equation}
S_{\text{NG}}=\int d^{2}\xi \sqrt{\det \partial _{\alpha }X^{a}\partial
_{\beta }X^{b}\eta _{ab}}.  \label{eq:ng}
\end{equation}%
where we have gone to tangent-space spacetime indices. We make the simple
observation that $\mathcal{L}$ is the square root of a homogeneous quartic
polynomial. In the case of a spacetime with signature $(2,2)$, it was shown in
\cite{Duff:2006uz} that $\mathcal{L}_{\text{NG}}$ is given by the square
root of Cayley's hyperdeterminant, a quartic invariant of $[\SL(2,\mathds{R}%
)]^{3}$,which is the quartic norm of the FTS $\mathfrak{F}(\mathds{R}\oplus %
\mathds{R}\oplus \mathds{R})$ appearing in the third row of \autoref%
{tab:FTSsummary}. Linearizing the quartic polynomial appearing in %
\eqref{eq:ng} we can write more generally
\begin{equation}
\mathcal{L}=\sqrt{|\Delta (\partial _{\alpha }X^{a},\partial _{\alpha
}X^{a},\partial _{\alpha }X^{a},\partial _{\alpha }X^{a})|}
\end{equation}%
where $\Delta $ is a totally symmetric quartic form defined by
\begin{equation}
\Delta (X_{\alpha _{1}}^{a_{1}},Y_{\alpha _{2}}^{a_{2}},Z_{\alpha
_{3}}^{a_{3}},W_{\alpha _{4}}^{a_{4}})=\frac{1}{12}\left[ \epsilon ^{\alpha
_{1}\alpha _{3}}\epsilon ^{\alpha _{2}\alpha _{4}}\eta _{a_{1}a_{2}}\eta
_{a_{3}a_{4}}+\text{5 $(\alpha _{i}a_{i})$ pairwise perms.}\right] X_{\alpha
_{1}}^{a_{1}}Y_{\alpha _{2}}^{a_{2}}Z_{\alpha _{3}}^{a_{3}}W_{\alpha
_{4}}^{a_{4}}.
\end{equation}%
The $6=4!/4$ distinct terms are generated by permutations modulo those of
the form $(ij)(kl)$ under which $\epsilon ^{\alpha _{i}\alpha _{j}}\epsilon
^{\alpha _{k}\alpha _{l}}\eta _{a_{i}a_{j}}\eta _{a_{k}a_{l}}$ is invariant.

Taking this as the definition of a  quartic norm allows us to define the Nambu-Goto Freudenthal triple system $\mathfrak{F}_{\text{NG}}$ by introducing a carefully chosen anti-symmetric bilinear form. This pair then defines the triple product which must be checked to satisfy the basic FTS identity \eqref{eq:def}.

Let us pick a basis for the string
world-sheet derivatives as a vector space,
\begin{equation}
F=F_{\alpha }^{a}e^{\alpha }\otimes e_{a},\qquad \text{where}\qquad
F_{\alpha }^{a}=\partial _{\alpha }X^{a},
\end{equation}%
and define,

\begin{enumerate}
\item The non-degenerate antisymmetric bilinear form (\textit{cfr.} \cite%
{Cerchiai:2009pi,Ferrara:2011di})
\begin{equation}
\{e^{\alpha }\otimes e_{a},e^{\beta }\otimes e_{b}\}=\epsilon ^{\alpha \beta
}\eta _{ab}=\Omega .
\end{equation}

\item The not identically zero symmetric quartic norm \cite{Ferrara:2011di} as determined by the  Nambu-Goto Lagrangian,
\begin{equation}\label{eq:K2}
\Delta (e^{\alpha }\otimes e_{a},e^{\beta }\otimes e_{b},e^{\gamma }\otimes
e_{c},e^{\delta }\otimes e_{d})=\frac{1}{12}\left[ \epsilon ^{\alpha \gamma
}\epsilon ^{\beta \delta }\eta _{ab}\eta _{cd}+\text{5 perms.}\right] =K.
\end{equation}
\end{enumerate}

Defining the triple product through
\begin{equation}
\{T(e^{\alpha }\otimes e_{a},e^{\beta }\otimes e_{b},e^{\gamma }\otimes
e_{c}),e^{\delta }\otimes e_{d}\}=2\Delta (e^{\alpha }\otimes e_{a},e^{\beta
}\otimes e_{b},e^{\gamma }\otimes e_{c},e^{\delta }\otimes e_{d})
\end{equation}%
one finds
\begin{equation}
T(e^{\alpha }\otimes e_{a},e^{\beta }\otimes e_{b},e^{\gamma }\otimes
e_{c})=T_{abc\delta }^{\alpha \beta \gamma d}e^{\delta }\otimes e_{d}
\end{equation}%
where
\begin{equation}
T_{abc\delta }^{\alpha \beta \gamma d}=K_{abcd^{\prime }}^{\alpha \beta
\gamma \delta ^{\prime }}\epsilon _{\delta ^{\prime }\delta }\eta
^{d^{\prime }d}=\frac{1}{12}\left[ \epsilon ^{\alpha
\gamma }\delta _{\delta }^{\beta }\eta _{ab}\delta _{c}^{d}+\text{5 perms.}%
\right] .
\end{equation}%
Using,
\begin{equation}
K_{abcd}^{\alpha \beta \gamma \delta }T_{a^{\prime }b^{\prime }c^{\prime
}\delta }^{\alpha ^{\prime }\beta ^{\prime }\gamma ^{\prime }d}=\frac{1}{12}%
\left[ K_{abcc^{\prime }}^{\alpha \beta \gamma \beta ^{\prime }}\epsilon
^{\alpha ^{\prime }\gamma ^{\prime }}\eta _{a^{\prime }b^{\prime }}+\text{5
perms.}\right]
\end{equation}%
one obtains
\begin{equation}
3\{T(X,X,Y),T(Y,Y,Y)\}=\frac{1}{6}\left[ K_{abcc^{\prime }}^{\alpha \beta
\gamma \beta ^{\prime }}\epsilon ^{\alpha ^{\prime }\gamma ^{\prime }}\eta
_{a^{\prime }b^{\prime }}+\text{5 perms.}\right] X_{\alpha }^{a}X_{\beta
}^{b}Y_{\gamma }^{c}Y_{\alpha ^{\prime }}^{a^{\prime }}Y_{\beta ^{\prime
}}^{b^{\prime }}Y_{\gamma ^{\prime }}^{c^{\prime }}
\end{equation}%
which, using \eqref{eq:K2}, gives
\begin{equation}
3\{T(X,X,Y),T(Y,Y,Y)\}=2\{X,Y\}\Delta (X,Y,Y,Y)
\end{equation}%
as required by the third postulate defining a Freudenthal triple system (%
\textit{cfr.} (\ref{eq:def})). Hence, the $2(s+t)$-dimensional vector space
spanned by $\{e^{\alpha }\otimes e_{a }\}$ forms a Freudenthal triple
system $\mathfrak{F}_{\text{NG}}$. The automorphism group is given by $\SL(2,%
\mathds{R})\times \SO(t,s)$ clearly corresponding to the Freudenthal triple
system over the semi-simple rank-3 Jordan algebra $\mathds{R}\oplus \mathbf{%
\Gamma }_{s-1,t-1}$ (of which the fourth and fifth rows in \autoref%
{tab:FTSsummary} are particular cases, see \cite{Borsten:2011nq} for details), namely:%
\begin{equation}
\mathfrak{F}_{\text{NG}}=\mathfrak{F}\left( \mathds{R}\oplus \mathbf{\Gamma }%
_{t-1,s-1}\right) .
\end{equation}%
In this context the $\SL(2,\mathds{R})$ factor is just a global subgroup of
the world-sheet diffeomorphisms.

Note, this FTS has previously appeared in physics literature  \cite{Gunaydin:2005zz,Gunaydin:2009dq}. In particular,   quasiconformal realisations over these FTS where used to capture the three dimensional U-duality groups as spectrum generating quasiconformal groups in \cite{Gunaydin:2009dq}.

Consequently, the Nambu-Goto world-sheet action may be written using the \~{F%
}-dual,
\begin{equation}
S_{\text{NG}}=\frac{1}{2}\int d^{2}\xi\{\tilde{F},F\}=\int d^{2}\xi\sqrt{%
|\Delta (F)|} \label{NGG}
\end{equation}%
It is now manifest that the Nambu-Goto action is in fact invariant under \~{F%
}-duality since,
\begin{equation}
\{\tilde{F},F\}\mapsto \{-F,\tilde{F}\}=\{\tilde{F},F\}.
\end{equation}%
The equations of motion and Bianchi identities are then simply,
\begin{equation}
d\star
\begin{pmatrix}
\tilde{F} \\
F%
\end{pmatrix}%
=0
\end{equation}%
where $\star $ denotes the world-sheet Hodge dual. Hence the equations of
motion and Bianchi identities are interchanged by \~{F}-duality.

Using the idea of the black hole potential we can introduce an equivalent Polyakov-type
action of the form,
\begin{equation}
S_{\text{pot}}=\frac{1}{2}\int d^{2}\xi\{F,\mathcal{S}(\Phi )F\}=-\frac{1}{2}\int d^{2}\xi F%
\mathcal{M}(\Phi )F=\int d^{2}\xi V_{\text{BH}}\left( \Phi ,F\right). \label{Spot}
\end{equation}%
Here $\mathcal{S}(\Phi )=\Omega \mathcal{M}(\Phi )=\epsilon ^{\alpha \beta
}M_{\beta \delta }(\varphi )\eta ^{ab}M_{bc}(\phi )\in \SL(2,\mathds{R}%
)\times \SO(t,s)$ is a function of the auxiliary scalar fields $\Phi :=(\varphi
,\phi )$, which parametrise the coset,
\begin{equation}
\frac{\SL(2,\mathds{R})}{\SO(2)}\times \frac{\SO(t,s)}{\SO(t)\times \SO(s)},
\end{equation}%
and satisfies $\mathcal{S}(\Phi )^2=-\mathds{1}, \mathcal{M}(\Phi )^T=\mathcal{M}(\Phi )$.

Then, one can then introduce the ``non-critical'' world-sheet \^{F}-duality as follows:
\begin{equation}
\hat{F}:=-\mathcal{S}(\Phi )F,\qquad \hat{\hat{F}}=-F.\label{deff}
\end{equation}%
Recalling (\ref{deff}), $S_{\text{pot}}$ can  be rewritten as
\begin{equation}
S_{\text{pot}}=\frac{1}{2}\int d^{2}\xi \{\hat{F},F\}, \label{Spot2}
\end{equation}
which makes    \^{F}-invariance manifest (\emph{cfr}.
 \autoref{sec:bh2}).

Let, $G=\hat{F}$ so that
\begin{equation}
d\star
\begin{pmatrix}
{F} \\
{G}%
\end{pmatrix}%
=0.  \label{eq:fe_bi}
\end{equation}%
This system of equations is invariant under $\GL(2(s+t),\mathds{R})$,
however one must also preserve the definition of $G$ and the invariance of
the equations of motion of any other fields present in the theory. As shown
in \cite{Cecotti:1988zz}, by following the analysis of Gaillard and Zumino
\cite{Gaillard:1981rj}, in $D=2$, and more generally in $D=4k+2$, the
maximal consistent duality group for $n$ $D/2$-form field strengths is $\SO%
(n,n)$. It is immediately apparent that the same is true for \eqref{eq:fe_bi} if we treat $F, G$ independently, as they would be in the presence of other interacting fields,   but where the spacetime indices play the role of $n$.

Note that by substituting the equations of motion for the $2+st$ auxiliary scalar fields $\Phi$ back into $S_{\text{pot}}$ (\ref{Spot2}), one obtains $S_{%
\text{NG}}$ (\ref{NGG}). These equations of motion are nothing but the criticality conditions for $V_{\text{BH}}\left(
\Phi ,F\right) $). Indeed, as discussed in section 2, for
groups \textquotedblleft of type $E_{7}$" the \textquotedblleft
non-critical" $\hat{F}$-duality reduces to the \textquotedblleft critical" $%
\tilde{F}$-duality, when evaluated at the critical points of $V_{\text{BH}}$%
.

\subsection{Beyond Nambu-Goto}\label{sec:E7string}

The Nambu-Goto action in $(t,s)$ signature has been described in terms of a specific class of FTS,
\begin{equation}
F=\partial_\alpha X^a e^\alpha\otimes e_a \in \mathfrak{F}_{\text{NG}}=\mathfrak{F}\left( \mathds{R}\oplus \mathbf{\Gamma }%
_{t-1,s-1}\right) ,
\end{equation}%
where
\begin{equation}
\int d^{2}\xi\sqrt{%
|\Delta (F)|}
=\int d^{2}\xi\sqrt{\det \partial _{\alpha }X^{a}\partial
_{\beta }X^{b}\eta _{ab}}.
\end{equation}
In this case the automorphism group, $\Aut(\FTS_{\text{NG}})=\SL(2, \R)\times\SO(t,s)$, has two factors which are naturally identified with the global subgroup of world-sheet diffeomorphisms and the spacetime Lorentz group, respectively.

However, formally we are free to consider the ``world-sheet'' action
\begin{equation}
S_\FTS=\int d^{2}\xi \sqrt{%
|\Delta (F(\tau, \sigma))|},
\end{equation}
for an arbitrary FTS, where $\FTS$ is an irreducible $\Aut(\FTS)$-module. In fact, recalling that Cartan's quartic $E_{7(7)}$ invariant $I_4$ reduces to  Cayley's hyperdeterminant in a canonical basis \cite{Kallosh:2006zs},  it was already suggested in \cite{Duff:2006ev} that $I_4$  provides a natural generalisation of the Nambu-Goto action in $(2,2)$ signature with an $\SO(6,6)$ Lorentz group  embedded in the larger $E_{7(7)}$ symmetry. This example corresponds to choosing the final row of \autoref{tab:FTSsummary}, given by $\FOs=\FTS(\JOs)$, the FTS defined over the Jordan algebra of split-octonionic $3\times 3$ Hermitian matrices.  Here   $F(\tau, \sigma)\in \FOs$ transforms as the irreducible fundamental $\rep{56}$ of $\Aut(\FOs)=E_{7(7)}$ and the quartic norm is given by Cartan's quartic invariant $\Delta(F)=I_4(F)$, defining an $E_{7(7)}$-invariant generalisation of the Nambu-Goto action,
\be
S_{E_{7(7)}}=\int d^{2}\xi \sqrt{%
|I_4 (F)|}=\frac{1}{2}\int d\tau d\sigma \{\tilde{F}, F\}. \label{SE7}
\ee
As before we can  write an equivalent Polyakov-type action (\ref{Spot2}) by using the ``non-critical'' world-sheet \^{F}-duality defined by (\ref{deff}), where in this case $\mathcal{S}(\Phi )=\Omega \mathcal{M}(\Phi )$ is a function of 70 auxiliary scalar fields parametrising the coset $E_{7(7)}/\SU(8)$.

As above, by substituting the equations of motion for the $70$ auxiliary scalar fields $\Phi$ back into $S_{\text{pot}}$ (\ref{Spot2}), one obtains $S_{E_{7(7)}}$ (\ref{SE7}).

When $\Aut(\FTS)$ is simple, as in this case,  what we had previously treated  as the world-sheet and spacetime  indices are unified in the irreducible representation carried by $\FTS$. To facilitate a string action interpretation a covariant split into world-sheet time/space derivatives must be made. Such a split is not necessarily unique, however, drawing on the conventional Nambu-Goto construction, a natural choice is given by the maximal embedding,
\be
E_{7(7)}\supset\SL(2,\R)\times\SO(6,6)
\ee
under which,
\be
\rep{56}\rightarrow \rep{(2,12)}+\rep{(1,32)}
\ee
so that
\be
F=(F_{\alpha}^{a}, \lambda^A)\qquad \text{where}\quad \alpha=1,2\quad a=1,\ldots,12\quad A=1,\ldots,32.
\ee
The $\rep{(2,12)}$ $F_{\alpha}^{a}$ admits the usual interpretation as the world-sheet derivatives of the target space embedding coordinates $F_{\alpha}^{a}=\partial_\alpha X^a$ for a string propagating in a spacetime with $(6,6)$ signature. On the other hand the $(\rep{1,32})$ $\lambda^A$ is a singlet under the $\SL(2, \R)$ and a spacetime Weyl spinor. Associating the $\SL(2, \R)$ indices with world-sheet derivatives, as we have done, implies $\lambda^A$ is an auxiliary field.

The $\SL(2, \R)$  representations only determine the transformation properties of the fields under the subgroup of global world-sheet diffeomorphisms. We still need to specify their transformation rules under local world-sheet reparametrisations. Identifying $F_{\alpha}^{a}$  with $\partial_\alpha X^a$ fixes it as a world-sheet 1-form as usual. This, coupled with the requirement of world-sheet diffeomorphism invariance,  implies that $\lambda^A$ transforms as a world-sheet scalar density of weight $(-1)$. This follows from the decomposition of $I_4(F)$ under $\SL(2,\R)\times\SO(6,6)$ which splits into three terms
\be\label{eq:I4decomp}
I_4(F) = \det(F_{\alpha}^{a}F_{\beta}^{b}\eta_{ab}) + \alpha \epsilon^{\alpha\beta}F_{\alpha a}F_{\beta b}[\Gamma^{ab}]_{AB}\lambda^A\lambda^B+\beta[\Gamma^{ab}]_{AB}\lambda^A\lambda^B[\Gamma_{ab}]_{A'B'}\lambda^{A'}\lambda^{B'}.
\ee
If $\sqrt{|I_4(F)|}$ is to transform homogeneously as a weight $(-2)$ scalar density as required for diffeomorphism invariance, then
each summand  in \eqref{eq:I4decomp} must individually transform as a world-sheet scalar density of weight $(-4)$. This  is true if and only if $\lambda^A$ is a scalar density of weight $(-1)$, as claimed. Note that, under the transformations generated by the rank-$4$ symmetric
para-quaternionic $64$-dimensional coset algebra $\mathfrak{e}_{7(7)}\ominus
(\mathfrak{sl}(2,\mathbb{R})\oplus \mathfrak{so}(6,6))$, the fields $%
F_{\alpha }^{a}$ and $\lambda ^{A}$ mix together; this is consistent with
the global $E_{7(7)}$ symmetry, which also fixes the real parameters $\alpha
$ and $\beta $ of (\ref{eq:I4decomp}).

It is here worth remarking that the Jordan algebras and FTS considered here have previously appeared in physics literature as the basis of generalised spacetimes  \cite{Gunaydin:2005zz}, but from a rather different perspective.

\section*{Acknowledgments}

SF would like to thank Paolo Aschieri and Armen Yeranyan for useful discussions. The work of LB is supported by an Imperial College Junior Research Fellowship. The work of MJD is supported by the STFC under rolling grant ST/G000743/1.  They are both grateful for conversations with Leo Hughes and Silvia Nagy. The work of SF is supported by the ERC Advanced Grant no. 226455 \textit{SUPERFIELDS}. LB is grateful for hospitality at the CERN theory division  (where he was supported by the above ERC Advanced Grant). The work of AM is supported in part by the FWO - Vlaanderen, Project No. G.0651.11, and in part by the
Interuniversity Attraction Poles Programme initiated by the Belgian Science Policy (P7/37).


\providecommand{\href}[2]{#2}\begingroup\raggedright\endgroup

\end{document}